# Coherent excitation of bound electron quantum state with quantum electron wavepackets

Du Ran[1,2,3*], Bin Zhang[1,*], Reuven Ianconescu[1,4], Aharon Friedman[5], Jacob Scheuer[1], Amnon Yariv[6], and Avraham Gover[1,†]

[1]School of Electrical Engineering - Physical Electronics, Center of Laser-Matter Interaction, Tel Aviv University, Ramat Aviv 69978, Israel

[2]School of Electronic Information Engineering, Yangtze Normal University, Chongqing 408100, China

[3]Fujian Key Laboratory of Quantum Information and Quantum Optics, Fuzhou University, Fuzhou 350116, China

[4]Shenkar College of Engineering and Design 12, Anna Frank St., Ramat Gan 5252626, Israel

[5]Schlesinger Family Accelerator Centre, Ariel University, Ariel 40700, Israel

[6]California Institute of Technology, Pasadena, California 91125, USA

† **Correspondence:**
Corresponding Author.
gover@eng.tau.ac.il;

* These authors have contributed equally to this work and share first authorship.

**Abstract**

We present a fully quantum model for the excitation of a bound electron based on the "free-electron bound-electron resonant interaction" (FEBERI) scheme. The bound electron is modeled as a quantum two-level system (TLS) at any initial quantum (qubit) state, and the free electron is presented as a pre-shaped quantum electron wavepacket (QEW). In the case that the QEW is short or modulated at optical frequency, the TLS quantum state may be coherently controlled with multiple modulation-correlated QEWs. For this case, we derive the transition probability of the TLS due to interaction with a multi-particle beam based on an analytical approximate solution of the Schrodinger equation that amounts to using Born's probabilistic interpretation of the quantum electron wavefunction. We verify the credibility of the analytical model at its validity ranges using a fully quantum density matrix computation procedure. It is shown that the transition probability can grow quadratically with the number of correlated QEWs, and theoretically - exhibit full Rabi oscillation. The study indicates a possibility of engineering the quantum state of a TLS by utilizing a beam of shaped QEWs.

# 1  Introduction

Recent technological advances enable the shaping of free electrons wavefunction in the transverse [1-3] and longitudinal [4-6] dimensions in an electron microscope setup. In the longitudinal propagation direction of the electron (or in time dimension), the energy and density expectation values of the electron wavefunction can be modulated by a coherent laser beam at optical frequencies [7-9], utilizing the scheme of photo-induced near-field electron microscopy (PINEM) [10]. The modulation is made possible by a multiphoton emission/absorption process in the near field of a nanostructure [7], a foil [11] or a laser-beat (pondermotive potential) [12,13]. After the PINEM interaction, the energy spectrum of a single quantum electron wavefunction (QEW) exhibits discrete energy sidebands spaced apart by the laser photon energy $\hbar\omega_b$. It was also shown that due to the nonlinear energy dispersion of electrons in free space drift, the discrete energy modulation of the QEW is converted into tight periodic density modulation (bunching) at atto-second short levels [14,15]. Furthermore, it was shown that single QEWs can be shaped and compressed to femtoseconds pulse duration by means of a chirping and streaking process with coherent THz or Infra-Red (IR) beams [16,17]. The reality of the finite size shape and the density modulation of the QEW in interaction with light was asserted and analyzed in the context of stimulated Smith-Purcell radiation [18,19]. It has been confirmed experimentally by acceleration/deceleration interaction of the pre-shaped/pre-modulated QEW with another coherent laser beam synchronous with the modulating laser [20,21]. The reality of the QEW sculpting and modulation features in stimulated radiative interaction and superradiance has been asserted also in the case of multiple modulation-correlated electron wavepackets [22,23], as extension of the classical case of a pre-bunched particle beam [18].

Shaping and transfer of coherence and quantum properties from light to free electrons wavefunction have received recently significant research attention within an emerging new research field of "quantum electron optics". It has been shown that the coherence and incoherence features of light, and even the quantum statistical states of light, can be transferred to the free electron quantum electron wavefunction by means of the PINEM process [24,25]. This paves the way for new applications of these emerging technologies for coherent control of quantum systems of light and matter using pre-shaped and optically modulated QEWs.

A simple example of coherent control of quantum states by electrons is the effect of free-electron bound-electron resonant interaction (FEBERI) proposed in Ref. [26]. In this process, a pre-shaped or pre-modulated beam of QEWs interacts with a bound electron modeled by a quantum two-level system (TLS). Such a TLS model is valid, because in general, in the linear response regime, the transition amplitude responds linearly to all possible transitions of quantum levels in matter with frequency-dependent amplitude. Therefore, targeting a single two-level transition does not reduce the generality of the model even if the targets have multiple levels. The QEWs induce excitations of the TLS when passing in the vicinity of the TLS target. It has been suggested that a beam probability-density modulation-correlated QEWs can interact resonantly with the TLS when harmonic $n$ of its optical frequency modulation matches the TLS quantum energy level transitions $n\hbar\omega_b = E_{2,1}$, where $\omega_b$ is the periodic temporal modulation frequency of the QEW density distribution and $E_{2,1} = E_2 - E_1$ is the quantum energy gap of the TLS. This assertion has raised a debate [27-29], but also a stream of numerous recently published papers relating to different aspects of this effect and its potential applications in electron microscopy, and in diagnostics and coherent control of qubits [30-33].

Using a fully quantum-mechanical analysis (both free and bound electrons quantized) of the FEBERI interaction with a single arbitrarily shaped QEW [34,35], we showed that the FEBERI effect can be applied for coherent control and interrogation of the qubit state of a target TLS. However, because the FEBERI effect is practically very weak for a single QEW and single TLS, it is necessary to consider interaction of the TLS with multiple modulation-correlated QEWs. Such a beam of density modulation-correlated QEWs can be produced if all electrons are energy modulated in the PINEM process by the same coherent laser beam and then develop the same phase-correlated bunching after drift to the PINEM interaction point. In this case, as proposed in the early paper of Gover and Yariv [32], based on a semi-classical model, coherent build-up of the quantum state of the TLS may be possible, including full Rabi oscillation between the TLS quantum levels. Such a process would be the analogue of spontaneous superradiance (in the sense of Dicke [36]) by a density modulated electron beam, which is characterized by emission rate proportional to $N^2$ where $N$ is the number of interacting electrons. This process is well established for bunched classical particle electron beam [18,37], but was also shown to take place when the electrons are density modulated on the level of quantum wavefunction (in expectation value) and modulation-correlated [22,38] Likewise, the semiclassical analysis results in quadratic growth $\propto N^2$ of the transition from the ground state to the upper state of the TLS at resonance condition [39].

In this paper, we study analytically the dynamics of a TLS excited by multiple finite size or modulation-correlated QEWs under the assumption that the interaction period is smaller than the relaxation time $T_1$ and the decoherence time $T_2$ of the TLS. The analytical model is based on a semi-classical approximation of the Schrodinger equation for the entangled free and bound electrons. We present an approximation in which the free electron quantum recoil is neglected. This amounts to using a probabilistic model for the free electron location, relying on Born's interpretation of the quantum electron wavefunction. Using a full quantum-mechanical density matrix computation of TLS state, we examine the applicability of the probabilistic approximation. A full quantum solution of the TLS and all interacting free electrons would involve consideration of their mutual entanglement via interaction through the TLS. This is not considered in the present article.    The paper is organized as follows: In section 2, we present the model and the theoretical framework of the FEBERI setup.

In sections 3 and 4, we apply the probabilistic model to interpret the excitation of a TLS with a single near point-particle (short size) QEWs and a beam of such QEWs respectively. Sections 5 and 6 present the excitation of a TLS with a single density modulated QEW and a beam of such modulation correlated QEWs. All approximate analytical expressions are compared to results of the quantum density matrix computations. Conclusions are presented in section 7.

## 2 System model and theoretical framework

The setup of our system model is shown in Fig.1, in which a thin free electron QEW propagates in proximity to a Hydrogen-like atom that is modeled as a TLS. The interaction of the free electron and bound electron is considered to be the Coulomb interaction. We denote the joint wavefunction of the free and bound electrons by $|\Psi(r,r',t)\rangle$. Then the dynamics of the considered system is governed by the Schrödinger equation,

$$i\hbar \frac{\partial |\Psi(r,r',t)\rangle}{\partial t} = (H_0 + H_I)|\Psi(r,r',t)\rangle, \qquad (1)$$

where $H_0 = H_{0F} + H_{0B}$ is the unperturbed Hamiltonian of the free electron and bound electron. $H_I$ represents the interaction Hamiltonian. In order to apply the analysis also to relativistic electrons, we use a "relativistic" Hamiltonian for the free electron of energy which we have derived by second order iterative approximation of Klein-Gordon equation, neglecting spin effect [41,48]:

$$H_{0F}(r) = \epsilon_0 + \boldsymbol{v}_0 \cdot (-i\hbar\nabla - \boldsymbol{p}_0) + \frac{1}{2\gamma_0^3 m}(-i\hbar\nabla - \boldsymbol{p}_0)^2. \qquad (2)$$

This corresponds to second order expansion of the relativistic energy dispersion of a free electron $E_p = \epsilon_0 + \boldsymbol{v}_0 \cdot (\boldsymbol{p} - \boldsymbol{p}_0) + (\boldsymbol{p} - \boldsymbol{p}_0)^2/2\gamma_0^3 m$, where $\epsilon_0 = \gamma_0 mc^2$ and $\boldsymbol{p}_0 = \gamma_0 m\boldsymbol{v}_0$. This "relativistic Hamiltonian" has been derived recently also directly from the Dirac equation [40] without the quadratic term.

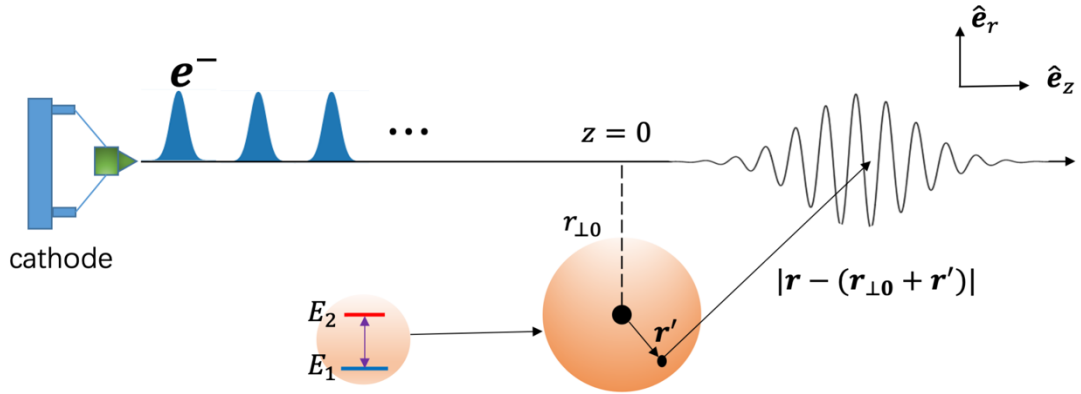

Fig. 1. Setup of free-electron bound-electron resonant interaction. A beam of multiple finite size or modulation-correlated QEWs interact with a bound electron which is modeled by a quantum two-level system.

For the simplified model where the spin is neglected, we assume that the free and bound electrons do not overlap spatially. Therefore, there are no exchange energy or spin–orbit interaction effects, and we can avoid the intricate second quantization of many-body interaction theory [41]. For the Coulomb interaction in the near field and neglecting retardation [42], the interaction Hamiltonian is

$$H_I(\boldsymbol{r}, \boldsymbol{r}') = \frac{e^2}{4\pi\varepsilon_0} \frac{\gamma}{[(\gamma z - z')^2 + (\boldsymbol{r}_{\perp 0} - \boldsymbol{r}'_\perp)^2]^{1/2}} \simeq \frac{e^2}{4\pi\varepsilon_0} \left[ \frac{1}{(\gamma^2 z^2 + r_{\perp 0}^2)^{1/2}} + \frac{\boldsymbol{r}' \cdot (\hat{\boldsymbol{e}}_z \gamma z - \hat{\boldsymbol{e}}_r r_{\perp 0})}{(\gamma^2 z^2 + r_{\perp 0}^2)^{3/2}} \right]. \quad (3)$$

Here we used Feynman's expression for the Coulomb potential [43] in order to keep the analysis valid in the relativistic regime. A more accurate form would be to use the Darwin potential for relativistic Coulomb interaction between moving charged particles [44,45]. However, for the parameters of the cases delineated here, the corrections due to this model are negligible.

Having the bound electron modelled as a TLS, the eigenfunction solutions of the Hamiltonian $H_{0B}$ satisfy

$$H_{0B}|\Psi_j(\boldsymbol{r}', t)\rangle = E_j|\Psi_j(\boldsymbol{r}', t)\rangle, \qquad (j = 1, 2) \quad (4)$$

where $|\Psi_j(\boldsymbol{r}', t)\rangle = \varphi_j(\boldsymbol{r}')e^{-iE_j t/\hbar}$. Then the general wavefunction of the bound electron can be represented as $|\Psi_B(\boldsymbol{r}', t)\rangle = \sum_{j=1}^{2} C_j |\Psi_j(\boldsymbol{r}', t)\rangle$, where $\sum_{j=1}^{2}|C_j|^2 = 1$. The wavefunction solution of the free electron in zero order is taken to be a general wavepacket

$$|\Psi_F^{(0)}(z, t)\rangle = \int \frac{dp}{\sqrt{2\pi\hbar}} c_p^{(0)} e^{-iE_p t/\hbar} e^{ipz/\hbar}, \quad (5)$$

The shape of the wavepacket is determined by the coefficients in momentum dimension - $c_p^{(0)}$. Here we consider two cases. The first case is a finite size QEW of Gaussian shape, arriving to the FEBERI interaction point $z = 0$ after a free drift length $L_d$ from its waist position. In 1-D momentum space it is presented as:

$$c_p^{(0)} = \frac{1}{\left(2\pi\sigma_{p_0}^2\right)^{1/4}} exp\left(\frac{-(p-p_0)^2}{4\sigma_{p_0}^2} - \frac{iE_p L_d}{\hbar v_0}\right), \tag{6}$$

The second case is a PINEM phase-modulated Gaussian wavepacket arriving to the FEBERI interaction point $z = 0$ after a free drift length $L_d$ from the PINEM modulation point:

$$c_{pM}^{(0)} = \frac{1}{\left(2\pi\sigma_{p_0}^2\right)^{1/4}} \sum_m J_m(2|g_L|) exp\left(\frac{-(p-p_0-\delta p_L)^2}{4\sigma_{p_0}^2} - im(\phi_0 + \omega_b t_0) - \frac{iE_p L_d}{\hbar v_0}\right), \tag{7}$$

Such a wavepacket represents in spatiotemporal space a density modulated finite size QEW [14]. Here $g_L$ is the PINEM coupling strength parameter, $\omega_b$ and $\phi_0$ are the frequency and reference phase of the laser-beam-induced field at the PINEM interaction point and $\sigma_{p_0}$ is the QEW momentum spread. For simplicity we assume that the QEW reaches the interaction point $z = 0$ at time $t_0$ at its longitudinal waist, so that the axial coordinate spread of the QEW is $\sigma_{z_0} = \frac{1}{2\hbar\sigma_{p_0}}$ (the expansion of the QEW during the short interaction time is negligible [19]).

During the interaction process, the expansion coefficients of the QEW $c_p^{(0)}$ get entangled with the coefficients of the bound electron $C_j$. Then the general combined wavefunction of the free electron and bound electron during the interaction can be represented in terms of the eigenfunctions:

$$|\Psi(r,r',t)\rangle = \sum_{j=1}^{2} \int dp\, c_{j,p}(t)\varphi_j(r') e^{-\frac{iE_j t}{\hbar}} c_p^{(0)} e^{-\frac{iE_p t}{\hbar}} e^{\frac{ipz}{\hbar}}. \tag{8}$$

After substituting this expression into Eq. (1) and cancelling out the no-interaction terms, we are left with

$$i\hbar \sum_{j=1}^{2} \int dp\, \dot{c}_{i,p}(t)\varphi(r') e^{-\frac{iE_j t}{\hbar}} e^{-\frac{iE_p t}{\hbar}} e^{\frac{ipz}{\hbar}} = H_I(r,r') \sum_{j=1}^{2} \int dp\, \dot{c}_{j,p}(t)\varphi(r') e^{-\frac{iE_j t}{\hbar}} e^{-\frac{iE_p t}{\hbar}} e^{\frac{ipz}{\hbar}} \tag{9}$$

By multiplying by $\varphi_i^*(r')$ and integrating over space, we reach an integro-differential equation that needs to be solved as function of time

$$i\hbar \int dp\, \dot{c}_{i,p}(t) e^{-\frac{iE_i t}{\hbar}} e^{-\frac{iE_p t}{\hbar}} e^{\frac{ipz}{\hbar}} = e^{-\frac{iE_j t}{\hbar}} \int dp\, c_{j\neq i,p}(t) M_{i,j}(r_{\perp 0}, r) e^{-\frac{iE_p t}{\hbar}} e^{\frac{ipz}{\hbar}}, \tag{10}$$

where $M_{i,j}(\boldsymbol{r}_{\perp 0},\boldsymbol{r}) = \langle i|H_I(\boldsymbol{r},\boldsymbol{r}')|j\rangle = \int d^3\boldsymbol{r}'\varphi_i^*(\boldsymbol{r}')H_I(\boldsymbol{r},\boldsymbol{r}')\varphi_j(\boldsymbol{r}')$, and we have used the ortho-normality relation $\int \varphi_i^*(\boldsymbol{r}')\varphi_j(\boldsymbol{r}')d^3\boldsymbol{r}' = \delta_{i,j}$ and defined the self-interaction term $\langle i|H_I(\boldsymbol{r},\boldsymbol{r}')|i\rangle = 0$. If $|\boldsymbol{r}'| \ll |\boldsymbol{r}-\boldsymbol{r}'| \simeq (r_{\perp 0}^2 + \gamma^2 z^2)^{1/2}$, the integration over $\boldsymbol{r}'$ in $M_{i,j}(\boldsymbol{r}_{\perp 0},\boldsymbol{r})$ can be carried out independently of $\boldsymbol{r}$, and for the interaction Hamiltonian (3) we have

$$M_{i,j}(\boldsymbol{r}_{\perp 0},\boldsymbol{r}) = -\frac{e}{4\pi\varepsilon_0}\frac{\boldsymbol{\mu}_{i,j}\cdot(\hat{\boldsymbol{e}}_z\gamma z - \hat{\boldsymbol{e}}_r r_{\perp 0})}{(\gamma^2 z^2 + r_{\perp 0}^2)^{3/2}}, \quad (11)$$

where $\boldsymbol{\mu}_{i,j} = -e\boldsymbol{r}_{i,j} = -e\int d^3\boldsymbol{r}'\varphi_i^*(\boldsymbol{r}')\boldsymbol{r}'\varphi_j(\boldsymbol{r}')$ is the dipole transition matrix element.

## 3 Probabilistic model for the excitation of TLS with single free electron

In order to describe the TLS dynamics analytically, we present in this section an iterative approach to solve the source equation (10). Substituting

$$c_{j,p}(t) \simeq C_j^{(0)}(t)c_p^{(0)}, \quad (12)$$

on the RHS of (10) allows calculation of the development in time of the TLS, neglecting the recoil dynamics of the QEW. Multiplying Eq. (10) by the complex conjugate of the free electron wavefunction Eq. (5) and integrating over space, one obtains

$$\frac{i}{2\pi}\int dp'\dot{c}_{i,p'}(t)c_p^{(0)*}e^{i(E_{p'}-E_p)t/\hbar}\int dz e^{i(p-p')z/\hbar} = C_j^{(0)}(t)e^{i\omega_{i,j}t}\int d^3\boldsymbol{r} M_{i,j}(\boldsymbol{r}_{\perp 0},\boldsymbol{r})\left|\Psi_F^{(0)}(\boldsymbol{r},t)\right|^2. \quad (13)$$

With $\int dz e^{i(p-p')z/\hbar} = 2\pi\hbar\delta(p-p')$, we have

$$2\pi i\hbar \int dp'\dot{c}_{i,p'}(t)c_{p'}^{(0)*} = C_j^{(0)}(t)e^{i\omega_{i,j}t}\int d^3\boldsymbol{r} M_{i,j}(\boldsymbol{r}_{\perp 0},\boldsymbol{r})\left|\Psi_F^{(0)}(\boldsymbol{r},t)\right|^2, \quad (14)$$

This presentation is reminiscent of interaction with an unperturbed point-particle that arrives at time $t_0$ at the interaction point $z = 0$ with Born's quantum wavefunction probability $\left|\Psi_F^{(0)}(\boldsymbol{r},t)\right|^2$. It should be stressed that $\left|\Psi_F^{(0)}(\boldsymbol{r},t)\right|^2$ is not well determined for a single electron. We assume that it is possible to solve Eq. (14) with substitution of its expectation value - $\langle\left|\Psi_F^{(0)}(\boldsymbol{r},t)\right|^2\rangle$, and the solution will then represent the result of interaction with an ensemble of identical QEWs.

The probability distribution of a single electron QEW of narrow width is:

$$\langle |\Psi_F^{(0)}(\mathbf{r},t)|^2 \rangle = \delta(r_{\perp 0}) f_{ez}(z - v_0(t-t_0)) = \delta(r_{\perp 0}) f_{et}(t - t_0 - z/v_0)/v_0, \quad (15)$$

where $f_{et}$ is normalized over time. Then Eq. (14) can be simplified to

$$i\hbar \int dp' \dot{c}_{i,p}(t) c_{p'}^{(0)*} = C_j^{(0)}(t) e^{i\omega_{i,j}t} f(t - t_0), \quad (16)$$

where $f(t - t_0) = \int dz\, M_{i,j}(z) f_{et}(t - t_0 - z/v_0)/v_0$ is the weighed interaction strength, which for a Gaussian can be calculated from the wavepacket (5). By neglecting the dynamics of the QEW around the interaction time $t_0$ also on the LHS, we can turn Eq. (16) into coupled differential equations for the TLS:

$$\dot{C}_i(t) = \frac{1}{i\hbar} C_j(t) e^{-i\omega_{i,j}t} f(t - t_0), \quad (17)$$

and after integration

$$C_i(t_0^+) = C_i(t_0^-) + \Delta C_i = C_i(t_0^-) + \frac{1}{i\hbar} \int_{t_0^-}^{t_0^+} dt\, C_j(t) e^{-i\omega_{i,j}t} f(t - t_0) \quad (18)$$

For a single Gaussian wavepacket at its longitudinal waist $\sigma_{z_0} = v_0 \sigma_{et}$, we have

$$f_{et}\left(t - t_0 - \frac{z}{v_0}\right) = \frac{1}{\sqrt{2\pi}\sigma_{et}} \exp\left(-\left(t - t_0 - \frac{z}{v_0}\right)^2 / 2\sigma_{et}^2\right). \quad (19)$$

Normalizing time to the transit time parameter $\bar{t} = t/t_r$, where $t_r = r_{\perp 0}/\gamma\beta c$, and defining $\bar{t}' = z/v_0 t_r$, the weighed interaction strength $f(t - t_0)$ can be recast into a convolution relation

$$f_{\parallel/\perp}(t - t_0) = \int dz\, M_{i,j}(z) f_{et}(t - t_0 - z/v_0)/v_0 = K_{\parallel/\perp} \int_{-\infty}^{+\infty} d\bar{t}' \frac{\bar{t}'}{(\bar{t}'^2 + 1)^{3/2}} \frac{1}{\sqrt{2\pi}\bar{\sigma}_{et}} e^{-(\bar{t} - \bar{t}_0 - \bar{t}')^2/2\bar{\sigma}_{et}^2}, \quad (20)$$

with the parameter $\bar{\sigma}_{et} = \sigma_{et}/t_r$ being the ratio of the wavepacket duration and the transit time and

$K_{\parallel/\perp} = e^2 \mathbf{r}_{i,j} \cdot \hat{\mathbf{e}}_{z/r}/4\pi\varepsilon_0 r_{\perp 0}^2$. Figure 2 shows the weighed interaction strength (20) between the TLS and free electron for the longitudinal and vertical dipole moment orientations with different electron wavepacket sizes quantified by parameter $\bar{\sigma}_{et}$. The maximum interaction strength decreases with the increasing of the wavepacket size. The longitudinal component of the electric field induced by the traversing electron at the position of the dipole reverses sign. This is the reason for the antisymmetric shape of the interaction strength as a function of time when the dipole moment is oriented longitudinally (Fig. 2(a)), and its symmetry when it is oriented vertically (Fig. 2(b)).

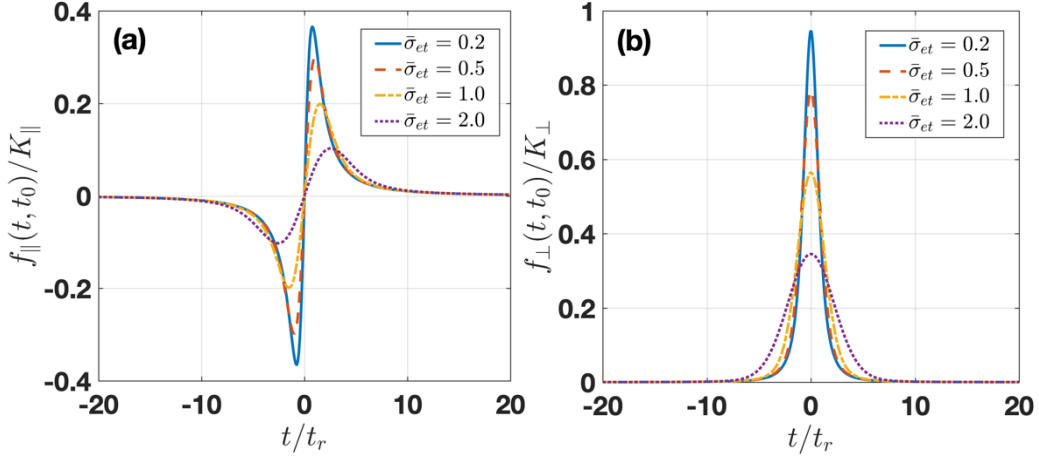

Fig. 2 The weighed interaction strength $f(t, t_0)$ for (a) $M_{i,j,\parallel}(z)$ and (b) $M_{i,j,\perp}(z)$.

Defining Fourier transform $F(\omega) = \mathcal{F}\{f(t-t_0)\} = \int_{-\infty}^{+\infty} dt\, e^{-i\omega t} f(t-t_0)$, the TLS transition amplitude during the interaction (17) turns out to be proportional to the Fourier transform of the weighted function (19) at the transition resonance frequency:

$$\Delta C_i = \frac{1}{i\hbar} C_j(t_0) F(-\omega_{i,j}), \qquad (21)$$

where

$$F(-\omega_{i,j}) = \frac{1}{v_0} \int_{-\infty}^{+\infty} dt\, e^{i\omega_{i,j}t} \int_{-\infty}^{+\infty} dz\, M_{i,j}(z) f_{et}\left(t - t_0 - \frac{z}{v_0}\right)$$

$$= \frac{1}{v_0} \int_{-\infty}^{+\infty} dz\, e^{i\omega_{i,j}\left(t_0 + \frac{z}{v_0}\right)} M_{i,j}(z) F_{et}(-\omega_{i,j})$$

$$= \frac{1}{v_0} e^{i\omega_{i,j}t_0} \widetilde{M}_{i,j}\left(\frac{\omega_{i,j}}{v_0}\right) F_{et}(-\omega_{i,j}). \qquad (22)$$

For a Gaussian QEW with $f_{et}(t-t_0) = \frac{1}{(2\pi\sigma_{et})^{1/2}} e^{-(t-t_0)^2/2\sigma_{et}}$, $F_{et}(-\omega_{i,j}) = e^{-\omega_{i,j}^2 \sigma_{et}^2/2}$. Then the incremental transition probability amplitude in Eq. (21) reads

$$\Delta C_i = \frac{1}{i\hbar v_0} C_j(t_0) e^{i\omega_{i,j}t_0} \widetilde{M}_{i,j}\left(\frac{\hbar\omega_{i,j}}{v_0}\right) e^{-\omega_{i,j}^2 \sigma_{et}^2/2}. \qquad (23)$$

Therefore, the transition probability of TLS after interaction is

$$P_i(t_0^+) = |C_i(t_0^-) + \Delta C_i|^2 = P_i^{(0)} + \Delta P_i^{(1)} + \Delta P_i^{(2)}, \qquad (24)$$

where $P_i^{(0)} = |C_i(t_0^-)|^2$, $\Delta P_i^{(1)} = 2Re[C_i(t_0^-) \cdot \Delta C_i]$, and $\Delta P_i^{(2)} = |\Delta C_i|^2$. For a finite size QEW, the transition probability to the upper-level quantum state for excitation of the TLS from its ground state ($C_1(t_0^-) = 1$ and $C_2(t_0^-) = 0$) is

$$P_2(t_0^+) = \Delta P_2^{(2)}(t_0^+) = \frac{1}{\hbar^2 v_0^2} \left|\widetilde{M}_{i,j}\left(\frac{\hbar\omega_{i,j}}{v_0}\right)\right|^2 e^{-\Gamma^2}, \qquad (25)$$

where $\Gamma = \omega_{i,j}\sigma_{et}$, and the first order term vanishes. In the case of excitation of the TLS from a superposition state, the second order transition term is the same as Eq. (25), while the first order transition term in Eq (24) is:

$$\Delta P_i^{(1)} = \frac{2}{\hbar v_0} \left|\widetilde{M}_{i,j}\left(\frac{\hbar\omega_{i,j}}{v_0}\right) C_i^{(0)*}(t_0) C_j^{(0)}(t_0)\right| e^{-\Gamma^2/2} \sin\zeta, \qquad (26)$$

where $\zeta = \phi - \omega_{2,1}t_0$ with $\phi$ being the phase of the quantum state of the TLS.

The probabilistic model approximation is presumed to apply in the very short QEW regime $\sigma_{et} \ll T_{2,1}$, corresponding to the limit of short interaction time $t_{int} < 1/\omega_{2,1}$. Eqs. (25, 26) manifest through the parameter $\Gamma$ the wavepacket size dependence of the transition probabilities for excitation of the TLS from ground state or superposition state in the near-point-particle parameters regime. Note that the excitation probability from a superposition state, the incremental transition probability $\Delta P_i^{(1)}$ also depends on the relative phase of the quantum state of the TLS. The phase-match timing dependence on the dipole oscillation phase at the short interaction impulse at the QEW arrival time to the interaction point, is indicative of a possible coherent interaction enhancement by multiple electrons with correlated arrival timing, as discussed in the following section.

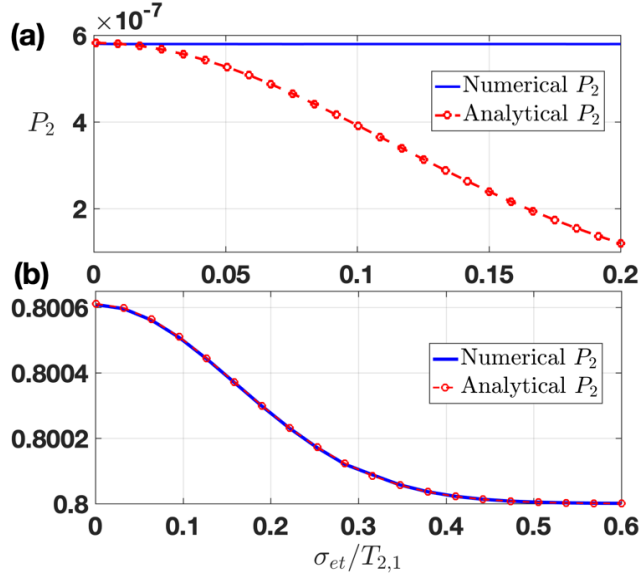

Fig. 3. Numerical solution of Schrodinger equation (1) and the analytical solution expressions of equation (24) for the transition probability dependence of the TLS on the QEW size: (a) for the initial TLS in the ground state and (b) for the initial TLS in a superposition state $|\psi_B\rangle_{in} = (|1\rangle + 2i|2\rangle)/\sqrt{5}$.

In order to check the validity of the analytical approximations, we have developed numerical computation codes for solving the FEBERI problem of interaction between a single finite size QEW and a TLS at any initial state, starting from the Schrödinger equation (1). The computation examples of the FEBERI effect were performed for a model of a Gaussian QEW and were studied as a function of its size $\sigma_{et}$, in order to examine the claimed dependence of the interaction on the wavepacket shape. In all the current examples the dipole polarization was taken to be transverse. The parameters used in the examples are typical to electron microscope PINEM-kind experiments [14], shown in Table 1. Note that the Hamiltonian of the combined system is time-independent, so that the solution of the Schrödinger equation (1) can be represented by the density matrix

$$\rho(t) = U(t)\rho(t_0)U(t), \quad (27)$$

where the evolution operator is $U(t) = \exp\left\{-\frac{i}{\hbar}(H_0 + H_I)t\right\}$ and the initial state of the combined system is $\rho(t_0) = |\Psi(t_0)\rangle\langle\Psi(t_0)|$. Because we are interested now in the excitation of the TLS, the state of the TLS can be obtained by tracing out the free electron states

$$\rho_b(t) = \text{Tr}_f[\rho(t)]. \quad (28)$$

The transition probability to the TLS upper quantum state is defined as $P_2 = \langle 2|\rho_b(t)|2\rangle$. It is then possible to generalize the simulation of the FEBERI to multiple QEWs, in which case a train of QEWs sequentially interacts with the bound electron. After each interaction of an electron with the TLS, we trace out the free electron states to find the state of TLS Eq. (28), which is used then as the initial state for the interaction with the next electron. We then repeat the calculation of Eq. (27). The computation algorithm details are given in the appendix of Ref. [39].

Table 1: The typical values of the simulation parameters.

| Physical parameters | Typical values |
| --- | --- |
| Beam Energy | $\epsilon_0 = 200 \; keV \; (\gamma_0 = 1.4)$ |
| Free electron impact parameter | $r_\perp = 2 \; nm$ |
| TLS energy gap (transition | $E_{2,1} = 2 \; eV \; (\omega_{2,1} = 3 \times 10^{15} \; rad/s)$ |
| Dipole moment | $\mu_{i,j} = 5 \; Debye$ |

Figure 3 displays the upper-level probability after interacting with a single QEW as a function of wave-packet size for different initial TLS quantum states. Numerical simulation results show that the transition probability is independent of the wave-packet size if the TLS starts from the ground state (blue curve), which contradicts the analytical result (Eq. 25) of exponential decay (red curve). This discrepancy is expectable, because our probabilistic model approximation, neglecting the free electron recoil, is presumed to apply only in the very short size QEW regime. Indeed, the quantum numerical computation result of finite wavepacket size-independent transition probability from ground state, seems to be more agreeable than the analytical result even with the philosophical point of view of Born's probability interpretation of the electron wavefunction: when $\sigma_{et}$ is large, the probability of the point-particle arrival to the TLS location is spread over a longer time, but it always happens at some time during the passage of the QEW, and must exhibit the same inelastic scattering, but the phase of the TLS dipole moment oscillation, undefined initially, is random after interaction. On the other hand, when the initial state of the coherently pre-excited TLS is a quantum superposition state of well-defined dipole oscillation phase, the quantum-mechanical numerical simulation result of the post-interaction probability is consistent with the analytical approximation expression of near point particle QEW with well-defined phase relative to the TLS dipole oscillation phase. In this case the transition probability strongly depends on the wavepacket size, decaying with the increase of the wave-packet size in either model.

## 4  Excitation of TLS with a bunched electron beam

The excitation of a TLS with multiple QEWs is theoretically an intricate multi-particle quantum interaction problem that involves entanglement of the free electron wavefunction with the states of TLS. Here, we resort again to the simple analytic approximate probabilistic model, in which we extend Eq. (15) to multiple particles

$$\left\langle\left|\Psi_F^{(0)}(\boldsymbol{r},t)\right|^2\right\rangle = \sum_{k=1}^{N}\left\langle\left|\Psi_k^{(0)}(\boldsymbol{r},t)\right|^2\right\rangle, \tag{29}$$

where $\left\langle\left|\Psi_k^{(0)}(\boldsymbol{r},t)\right|^2\right\rangle = \delta(r_{\perp 0})f_{et}(t - t_{0k} - z/v_0)$. We then solve for the cumulative incremental transition probability for the case of periodically injected near-point-particle QEWs. Under the assumption that the relaxation time of the TLS is much longer than the duration of the $N$ QEWs pulse, we substitute the $N$ particles probability function $f(t - t_0) = \sum_{k=1}^{N}\frac{1}{v_0}\int dz M_{i,j}(z)f_{et}(t - t_0 - z/v_0)$ in Eq. (18), and changing order of integration in $z$ and $t$ results in

$$C_i(t_{0N}^+) = C_i(t_0^-) + \frac{1}{i\hbar v_0}\int_{t_0^-}^{t_0^+} dz M_{i,j}(z) \sum_{k=1}^{N} C_j(t_{0k}) \int dt\, e^{-i\omega_{i,j}t} f_{et}(t - t_{0k} - z/v_0) \tag{30}$$

With change of variable $t' = t - z/v_0$,

$$\begin{aligned}C_i(t_{0N}^+) &= C_i(t_0^-) + \frac{1}{i\hbar v_0}\int_{t_0^-}^{t_0^+} dz M_{i,j}(z) e^{\frac{i\omega_{i,j}z}{v_0}}\sum_{k=1}^{N} C_j(t_{0k}) \int dt'\, e^{-i\omega_{i,j}t'} f_{et}(t' - t_{0k})\\ &= C_i(t_0^-) + \frac{1}{i\hbar v_0}\widetilde{M}_{i,j}\left(\frac{\omega_{i,j}}{v_0}\right)\sum_{k=1}^{N} C_j(t_{0k})\, e^{i\omega_{i,j}t_{0k}} F_{et}(\omega_{i,j}).\end{aligned} \tag{31}$$

The incremental probability amplitude in this equation averages to zero for random $t_{0k}$, except when $\omega_{i,j} = n\omega_b$, where $t_{0k} = 2\pi k/\omega_b$, namely, when the QEWs arrive to the interaction point at a rate that is a sub-harmonic of the transition frequency $\omega_{i,j}$. Then, with the approximation of small change in the amplitude $C_1(t_{0k}) \cong C_1(t_0^-) = 1$, the amplitude of the upper level is

$$C_2(t_{0N}^+)\big|_{\omega_{i,j}=n\omega_b} \cong N\frac{1}{i\hbar v_0}\widetilde{M}_{2,1}\left(\frac{\omega_{2,1}}{v_0}\right) F_{et}(\omega_{2,1}). \tag{32}$$

For a Gaussian QEW (19), the transition probability to upper level for the $N$ QEWs case is

$$P_2(t_{0N}^+) = N^2\left\{\frac{1}{\hbar v_0}\left|\widetilde{M}_{2,1}\left(\frac{\omega_{2,1}}{v_0}\right)\right|\right\}^2 e^{-\omega_{i,j}^2\sigma_{et}^2}. \tag{33}$$

As we learned from the previous chapter, this approximate result may not be rigorous in the initial stage of the multiple electrons transition buildup from ground state, when the phase of the dipole moment oscillation is not well-defined. We conject that when $N$ is large enough, the phase of the TLS gets established by the first near-point-particle QEWs of the train, and the subsequent QEWs then continue to build up the transitions in phase.

This case of a periodically spaced train of near-point-particle QEWs may be realistic for low (microwave or THz) frequency TLS transitions, where classical Klystron-kind electron current modulation is available. It has thus been termed a "Quantum Klystron" in [30]. It can be comprehended as the quadratic approximation of the $\sin^2(\Omega_R t/2)$ scaling of a Rabi oscillation process with Rabi frequency $\Omega_R$, and it is the analogue of the classical bunched-particles beam superradiance effect [18] Note that in the classical point particle limit and low (microwave) frequencies [30] high current density of the electron beam is allowed (with the limitations of beam quality and space charge effect) and there may be then multiple electrons per period. We also point out that the case of a multiple periodic train of QEWs, is closely related to the earlier studied effect of "pulsed beam scattering" [46,47].

It is instructive to compare the quadratic dependence of (33) on the number of QEWs - $N^2$ to the same dependence in the case of superradiance [18,23]. In this comparison, the exponential decay factor $e^{-\Gamma^2}$ that originates from the finite size of the Gaussian QEW (19), is the quantum limit of the "bunching coefficient" in a bunched point-particle beam superradiance. In the case of a long wavepacket, the non-decaying probability expression for $P_2$ becomes relevant, and it would result in a multiple electrons TLS transition rate proportional to $N$, in analogy to the "shot-noise" spontaneous radiation emission limit of bunched-beam superradiance in the large bunching coefficient limit.

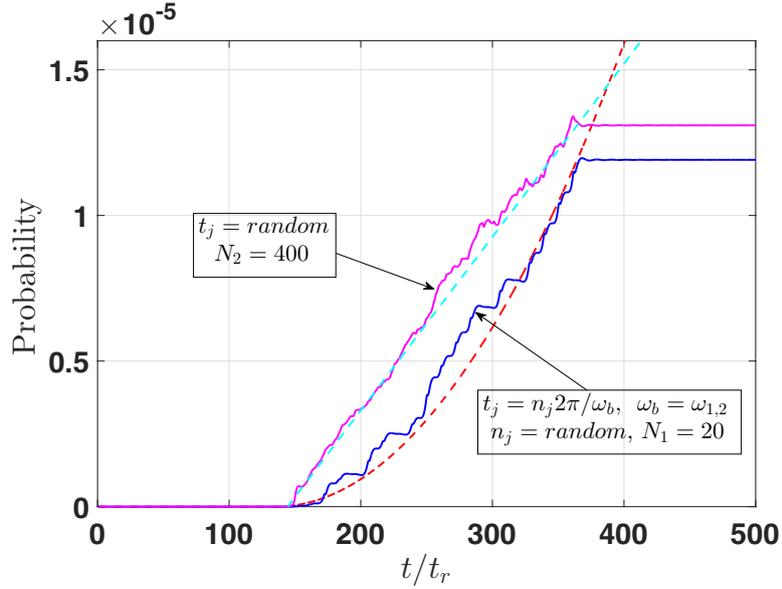

Fig. 4. Simulation results of the TLS upper-level probability buildup by electrons arriving to the interaction point at random (magenta) and by electrons arriving in-phase with the modulating laser (modulo the bunching period $T_b = 2\pi/\omega_b$) at the resonance condition $\omega_{2,1} = n\omega_b$ (blue). The light blue-dash and red-dash curves are the linear and quadratic curve-fittings, respectively.

Figure 4 shows simulation of the buildup of the TLS upper-level probability with $N_1 = 20$ particles arriving at times $t_j = t_{0j} + n_j T_b$, where $n_j$ is a random number. For the electrons arriving to the interaction point in-phase with the modulating laser modulo the bunching period $T_b = T_{2,1}$, the probability growth is evidently quadratic, $P_2 \propto N^2$ as claimed. For comparison, we show in the figure by the magenta curve the case where $t_j$ is taken to be entirely random. The growth rate is linear and the upper-level probability arriving at the same value needs $N_2 = N_1^2 = 400$ particles [29].

## 5 Excitation of a TLS with a modulated single free quantum electron wavepacket

Here we extend our Born's probability interpretation of the electron wavefunction to model the case of a density modulated QEW. In Eq. (29) we model the density expectation value of a single electron wavepacket as

$$\left\langle \left| \Psi_F^{(0)}(\mathbf{r},t) \right|^2 \right\rangle = \delta(r_{\perp 0}) f_{et}\left(t - t_0 - \frac{z}{v_0}\right) f_{mod}(t - \frac{z}{v_0} - t_L), \qquad (34)$$

where $f_{et}\left(t - t_0 - \frac{z}{v_0}\right)$ is the envelope function of the drifting QEW and the modulation function is periodic, and therefore can be expressed in terms of a Fourier serries $f_{mod}(t) = \sum_{m=-\infty}^{\infty} f_m e^{-im\omega_b t}$. The coefficients $f_m$ were derived in Ref. [22] for the case of the wavefunction of a modulated Gaussian QEW, $\omega_b t$ is the modulation phase, determines by the modulating laser beam.

We next derive the incremental excitation probabilities in Eq. (24). For a modulated QEW, the weighed probability function becomes

$$f(t - t_0) = \frac{1}{v_0} \int dz\, M_{i,j}(z) f_{et}(t - t_0 - z/v_0) \sum_{m=-\infty}^{\infty} f_m e^{i\omega_b(t - \frac{z}{v_0} - t_L)}. \quad (35)$$

We substitute this probability distribution of a modulated QEW in Eq. (18) and change the order of integrations

$$C_i(t_0^+) = C_i(t_0^-) + \frac{1}{i\hbar v_0} \int dz M_{i,j}(z) C_j(t_0) \sum_{m=-\infty}^{\infty} \int dt\, e^{-i\omega_{i,j}t} f_{et}(t - t_0 - z/v_0) f_m e^{i\omega_b(t - \frac{z}{v_0} - t_L)}. \quad (36)$$

With change of variables $t' = t - z/v_0$,

$$C_i(t_0^+) = C_i(t_0^-) + \frac{1}{i\hbar v_0} \widetilde{M}_{i,j}\left(\frac{\omega_{i,j}}{v_0}\right) C_j(t_0) \sum_{m=-\infty}^{\infty} f_m e^{-i(\omega_{i,j} - m\omega_b)t_0} F_{et}(\omega_{i,j} - m\omega_b) e^{im\omega_b t_L}. \quad (37)$$

Then the first-order incremental transition probability is

$$\Delta P_i^{(1)}(t_0^+) = \text{Re}\left\{\frac{1}{i\hbar v_0} \widetilde{M}_{i,j}\left(\frac{\omega_{i,j}}{v_0}\right) C_i^{(0)*}(t_0) C_j^{(0)}(t_0) \sum_{m=-\infty}^{\infty} f_m e^{i(\omega_{i,j} - m\omega_b)t_0} F_{et}(\omega_{i,j} - m\omega_b) e^{-im\omega_b t_L}\right\}. \quad (38)$$

If the envelope Gaussian distribution is a wide function -$\sigma_{et} > 2\pi/\omega_b$, then the spectral function $F_{et}(\omega_{i,j} - m\omega_b) = \exp\{-(\omega_{i,j} - m\omega_b)^2 \sigma_{et}^2/2\}$ is a narrow function around a harmonic $m = n$ that is resonant with the transition frequency $\omega_{i,j} = n\omega_b$. In such case, only one harmonic - $n$ can excite resonantly the transition:

$$\Delta P_i^{(1)}(t_0^+) = \text{Re}\left\{\frac{1}{i\hbar v_0} \widetilde{M}_{i,j}\left(\frac{\omega_{i,j}}{v_0}\right) C_i^{(0)*}(t_0) C_j^{(0)}(t_0) f_n e^{i(\omega_{i,j} - n\omega_b)t_0} e^{-in\omega_b t_L} e^{-(\omega_{i,j} - n\omega_b)^2 \sigma_{et}^2/2}\right\}, \quad (39)$$

still under the condition that it is phase-matched to the phase of the initial dipole moment oscillation phase $C_i^{(0)*}(t_0^-) C_j^{(0)}(t_0^-)$.

Likewise, the second-order incremental transition probability can be calculated

$$\Delta P_i^{(2)}(t_0^+) = \left|\frac{1}{\hbar v_0} \widetilde{M}_{i,j}\left(\frac{\omega_{i,j}}{v_0}\right) C_j^{(0)}(t_0)\right|^2 |f_n|^2 e^{-(\omega_{i,j} - n\omega_b)^2 \sigma_{et}^2}. \quad (40)$$

It can be seen that both first order and second order expressions of the incremental probabilities are dependent on the QEW shape and modulation features and display resonant excitation characteristics around the condition $\omega_{i,j} = n\omega_b$, which would manifest the QEW modulation characteristics in a properly set experiment. Note that in a modulated QEW the QEW envelop is necessarily longer than the modulation period: $\omega_{i,j}\sigma_{et} > 1$. Comparing the first and second order transition probability expressions of the modulated QEW (39,40) to the corresponding terms of the unmodulated finite size QEW (25, 26), we find out that at this limit the latter decay to zero, but the former (modulated QEW expressions) do not decay, as long as a harmonic of the modulation frequency is synchronous with the transition frequency $-(\omega_{i,j} - n\omega_b)\sigma_{et} \ll \pi$. This indicates a possibility for measuring the modulation features of the QEW. However, note that for single modulated QEWs, there is no enhancement of the transition probability even at resonance. At resonance, Eqs. (39,40) reduce to Eqs. (25, 26), except for a Fourier serries component coefficient.

We check the results of the probabilistic model approximation by comparing it to the results of the quantum density matrix model numerical computation that is based on solution of Schrodinger equation (1), as described in the appendix of Ref. [39]. While the analytical expressions provide only post-interaction transition probabilities, the numerical computation, interestingly enough, lets us follow the dynamics of the TLS quantum transition probability during the interaction time (the QEW transit time at proximity to the TLS) and the final transition probability after passage. Figures 5b and 5c display computation results of the dynamics of quantum transition to the upper state of the TLS for two cases of pre-shaped QEW distributions (see Fig. 5a): an unmodulated (broad) QEW ($\frac{\sigma_{et}}{T_{2,1}} = \frac{\omega_{2,1}\sigma_{et}}{2\pi} = 1$) (red curve) and a modulated QEW with the same size envelope ( $\sigma_{et}/T_{2,1} = 1$) (blue curve). The density modulation was pre-evaluated assuming a PINEM process [20] with the beam parameters of Table 1 and $g_L = 0.75$.

Figs. (5b) and (5c) present the dynamic buildup of the transition probability starting from ground state $|\Psi_B\rangle_{t_{in}} = |1\rangle$ and from a coherent superposition (qubit) state $|\Psi_B\rangle_{in} = \sin\frac{3\pi}{8}|1\rangle + \cos\frac{3\pi}{8}|2\rangle$, correspondingly. The bunching frequency was set to synchronize with the transition frequency at the fundamental harmonic $\omega_b = \omega_{2,1}$. The dynamics of transitions with the modulated QEW suggest gradual "quantum jumps" any time a sub-bunch of the QEW arrives to the FEBERI interaction point. This seems to be consistent with the Born picture of probability of point particle arrival. However, these are only probabilities of interaction events, the measurable post-interaction transition probability of the TLS from ground state (Fig. 5b) generated by a single passing-by resonantly modulated QEW is the same as with a single unmodulated near-point-particle QEW. This is in partial agreement with Eqs. (26,40) except for the exponential decay factors in these two expressions that should be set equal 1 because

the analytical model, neglecting quantum recoil, is not rigorous for a long QEW (see also blue curve in Fig. 3a). On the other hand, when starting from a coherent composition state of the TLS (Fig. 5c) the computation confirms the analytical expressions (39) (for modulated QEW at resonance) and (26) (for unmodulated long QEW): high transition probability in the first case and diminished transition probability in the latter (note that in this case the simulation result reflects practically only the contribution of the dominant first order incremental probability terms, because the second order contributions are minute).

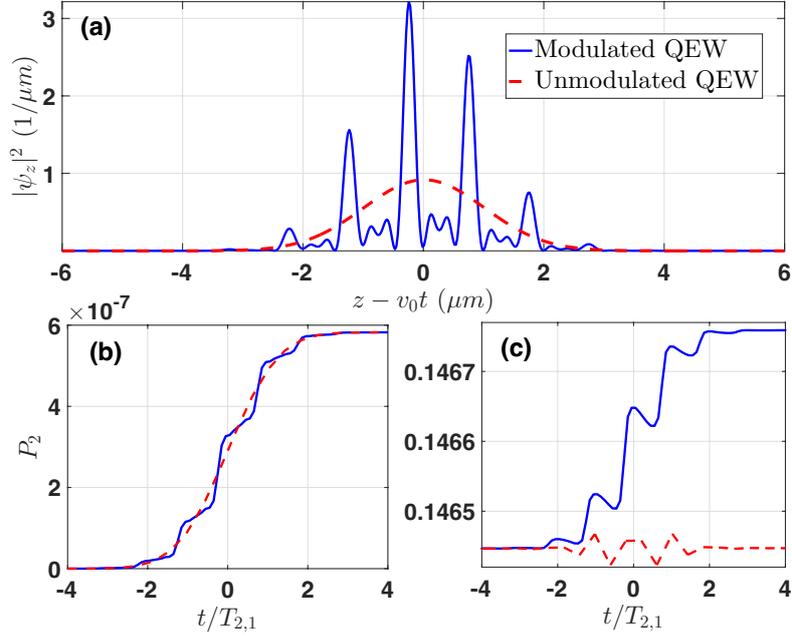

Fig. 5. (a) Density distribution of a broad QEW ($\sigma_{et}/T_{2,1} = 1$) (red broken line) and a modulated QEW having the same size envelope (blue continuous line). (b) Numerical computation results of the dynamics of the FEBERI interaction transition probability $P_2$ starting form ground state, based on the quantum model (solving the Schrodinger equation (1)) for the corresponding unmodulated and modulated QEWs. (c) The computed dynamics of the FEBERI interaction incremental transition probabilities for the same unmodulated and modulated QEWs, starting from a coherent superposition state of the TLS $|\Psi_B\rangle_{in} = \sin\frac{3\pi}{8}|1\rangle + \cos\frac{3\pi}{8}|2\rangle$.

## 6  Excitation of a TLS with a beam of modulation-correlated quantum electron wavepackets

In this section, we extend our Born's probability interpretation analytical model to the case of multiple modulation-correlated QEWs. Consider the case of multiple long size QEWs, all phase (energy) modulated at a PINEM interaction point at the level of their quantum wavefunctions [14] by the same coherent laser beam of frequency $\omega_b$ and phase $\omega_b t_L$. Passing all the same drift length to the FEBERI interaction point, the expectation values of their density are modulation phase correlated. We extend the probability distribution of a single electron QEW Eq. (15) to the modulation-correlated multiple particles by substitution in Eq. (29)

$$\left\langle \left|\Psi_k^{(0)}(\mathbf{r},t)\right|^2 \right\rangle = \delta(r_{\perp 0}) f_{et}\left(t - t_{0k} - \frac{z}{v_0}\right) f_{mod}(t - \frac{z}{v_0} - t_L), \qquad (41)$$

where $t_{0k}$ are the centroid arrival times of the envelopes of the modulated QEWs, and the modulation function, common to all QEWs is periodic in time $f_{mod}(t) = f_{mod}(t + 2\pi/\omega_b)$. For modulation-correlated QEWs, the weighed probability distribution function is

$$f(t - t_0) = \frac{1}{v_0} \sum_{k=1}^{N} \int dz\, M_{i,j}(z) f_{et}(t - t_{0k} - z/v_0) \sum_{m=-\infty}^{\infty} f_m e^{i\omega_b(t - \frac{z}{v_0} - t_L)}. \qquad (42)$$

Then, substitution in Eq. (18), changing the integration order of $z$ and $t$, results in

$$C_i(t_{0N}^+) = C_i(t_0^-) + \frac{1}{i\hbar v_0} \widetilde{M}_{i,j}\left(\frac{\omega_{i,j}}{v_0}\right) \sum_{k=1}^{N} \sum_{m=-\infty}^{\infty} C_j(t_{0k}) f_m e^{i(\omega_{i,j} - m\omega_b)t_{0k}} F_{et}(\omega_{i,j} - m\omega_b) e^{-im\omega_b t_L}. \qquad (43)$$

Again, for a broad Gaussian distribution $\sigma_{et} > 2\pi/\omega_b$, the spectral function $F_{et}(\omega_{i,j} - m\omega_b) = exp\{-(\omega_{i,j} - m\omega_b)^2 \sigma_{et}^2/2\}$ is a narrow function around a harmonic $m = n$ that is resonant at the transition frequency $\omega_{i,j} = n\omega_b$. Take $i = 2, j = 1$ (upper and lower levels), then with the approximation of small change in the amplitude $C_1(t_{0k}) \cong C_1(t_0^-) = 1$, we have

$$C_2(t_{0N}^+) \cong \frac{1}{i\hbar v_0} \widetilde{M}_{2,1}\left(\frac{\omega_{2,1}}{v_0}\right) \sum_{k=1}^{N} \sum_{m=-\infty}^{\infty} f_m e^{i(\omega_{2,1} - m\omega_b)t_{0k}} e^{-(\omega_{2,1} - m\omega_b)^2 \sigma_{et}^2/2} e^{-im\omega_b t_L}. \qquad (44)$$

This averages to zero for random arrival times $t_{0k}$ of the wavepacket centroids, except at the resonance case $\omega_{i,j} = m\omega_b$, where

$$C_2(t_{0N}^+)|_{\omega_{i,j}=m\omega_L} \cong \frac{N}{i\hbar v_0} \widetilde{M}_{2,1}\left(\frac{\omega_{2,1}}{v_0}\right) f_n e^{-im\omega_b t_L}. \qquad (45)$$

And independently of the arrival times $t_{0k}$, the transition probability to the upper level is

$$P_2(t_{0N}^+) = N^2 \left\{ \frac{1}{\hbar v_0} \left| \widetilde{M}_{2,1}\left(\frac{\omega_{2,1}}{v_0}\right) f_n \right| \right\}^2. \tag{46}$$

This expression explicitly manifests the $N^2$ scaling buildup of the upper quantum level probability in the case of multiple modulation-correlated QEWs, similar to the case of periodically modulated point particles and in analogy to superradiance of bunched particles [18].

To check the result of the probabilistic model approximation, we compare it to the results of the quantum model (Schrodinger equation) numerical computations described in chapter 3 [see details appendix of Ref. [39]. For simulating the dynamics of multiple QEWs in interaction with the TLS, we consider a train of incoming QEWs, sequentially interacting with the TLS, and repeat the calculation for each electron passing by. Equation 7 is used to represent the initial quantum electron wavefunction of each electron upon arrival to the FEBERI interaction point. In Eq. 7 we use a random value $t_{0j}$ (replacing $t_0$) corresponding to random arrival of the QEWs envelope centroids to the interaction point. The initial state of the TLS in each interaction is obtained by tracing out the free electron states in the post-interaction density matrix of the prior interaction. Possible overlap of QEWs is neglected, assuming a sparce electron beam. In Fig. 6, we show the results of the simulation for the parameters given in Table 1. The modulation density of each QEW in spatiotemporal domain is shown in Fig. 6(a). The wavepacket envelope size is $\sigma_{et}/T_{2,1} = 1$. The simulation results, depicted in Fig. 6 display the quadratic buildup (blue curve) of the upper quantum level probability in the case of multiple modulation-correlated QEWs at resonant fundamental harmonic FEBERI interaction condition $\omega_b = \omega_{2,1}$. Even though the QEW centroids arrive at random times $t_{0j}$, the phase $\phi_0$ (in Eq. 7) of all QEWs is the same, determined by the phase of the coherent laser beam that pre-modulates them all by a PINEM process. The noisy linear buildup curve (red) corresponds to the case of PINEM interaction with an incoherent light, where the phase $\phi_{0j}$ (replacing $\phi_0$ in Eq. 7) is random, as well as the envelop centroid arrival times $t_{0j}$.

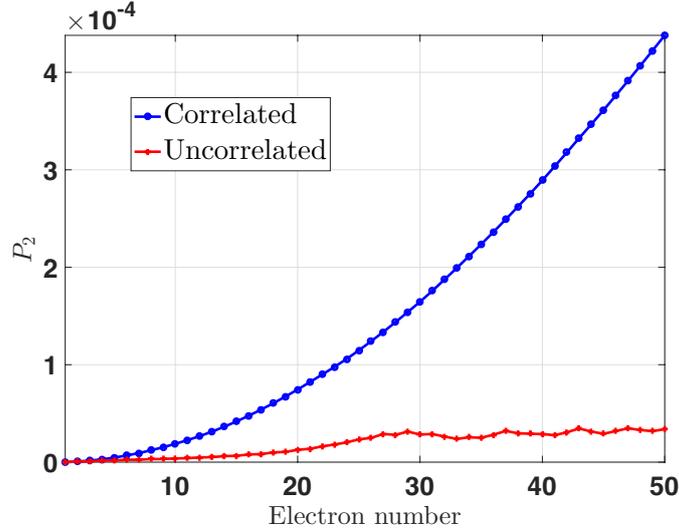

Fig. 6. Numerical simulation results of the transition probability $P_2$ based on the quantum density matrix model (solution of the Schrodinger equation (1)) in the cases of modulation-correlated and random QEW beams. In the first case (modulation correlated QEWs), the centroids of the QEWs arrive to the FEBERI interaction point at random time $t_{0j}$, but the modulation phase $\phi_0$ of all of them is the same. This results in quadratic scaling of the TLS excitation build-up (blue curve). In the second case (entirely random beam of modulated QEWs), the scaling of the TLS excitation build-up is a noisy linear curve (red). In comparing these curves to the semiclassical model results of Fig. 4, note that the horizontal axis here represents electrons number rather than time.

## 7      Conclusions

We studied the excitation of a bound electron in an arbitrary initial quantum (qubit) state by a free electron beam, where the bound electron is modeled as a quantum two-level system (TLS) and the free electrons are represented as quantum electron wavepackets (QEWs). We developed an analytical model for the FEBERI interaction based on Born's probabilistic interpretation of the quantum electron wavefunction and derived expressions for the transition probability of the TLS subject to interaction with a shaped QEW (finite size gaussian), optically modulated QEWs, and a beam of such QEWs. We tested the analytical results against the simulation results of an accurate quantum model of the FEBERI effect, based on density matrix solution of Schrodinger equation for the entangled free-bound electron wavefunctions.

   The accurate quantum model simulations show quadratic growth of the TLS quantum transition probability as a function of the number of electrons in a train of modulation-correlated QEWs, when an harmonic of the QEW modulation frequency is resonant with the TLS quantum transition frequency, and the modulation initial phase of all electrons is the same (*e.g.* they are modulated by the same coherent laser beam in a PINEM process). This result confirms the prediction of the analytical approximation model based on Born's probabilistic interpretation of the electron wavefunction and the earlier semiclassical analysis of the FEBERI effect [26].

The quadratic scaling of the TLS excitation build-up in the FEBERI process can be explained as the second order expansion of the sinus-squared scaling of a Rabi oscillation process [26], where the optical frequency near-field of the beam of modulation-correlated QEWs play the same role as a laser beam in a conventional Rabi oscillation experiment. Another instructive observation that came out of the numerical simulations is that in the case of modulation-correlated QEWs beam, the temporal intervals between the probability-density microbunches of the QEWs (Fig. 5) are spaced apart at resonance an integral number of times the TLS transition period ($T_{2,1} = 2\pi/\omega_{2,1}$) even though the centroids of the QEWs arrive at random. The coherent quadratic buildup can be then viewed in Born's probabilistic interpretation as a result of quasi-periodic arrival times of a train of quantum- probability-determined "point-particles" in-phase with the TLS dipole moment oscillation at the quantum transition frequency. This is a bridge to the semiclassical case of FEBERI interaction with a point-particle density-modulation beam (quantum klystron [30]) and an analogue of the radiative process of bunched electron beam superradiance [18].

To test the surprising result of coherent transition probability buildup with a modulation-correlated electron beam independently of the random arrival times of the electrons to the FEBERI point, we show in Fig. 6 (red curve) simulation of interaction with a beam of modulated QEWs that are not modulation phase-correlated (random $\phi_{0j}$). In this case, that corresponds to electron beam PINEM modulation with an incoherent laser beam, the FEBERI scaling is a noisy linear curve, indicating uncorrelated excitation of the TLS. This observation is in line with recent observations that PINEM modulation carries the coherence properties and even the quantum statistical fingerprint of the modulating light beam, and transfers it to cathodoluminescence spectrum [25] and to the EELS spectrum [24].

We point out that coherent control and demonstration of Rabi oscillation of a single TLS with the proposed FEBERI effect is attractive in comparison to such operations with a laser beam, because of the atomic scale resolution of an electron beam. However, the experimental realization of this scheme is challenging due to the very small value of the interaction coupling factor in typical TLS targets ($g \approx 10^{-3}$). Enhancement of the interaction may be possible in consideration of high dipole moment targets, such as multiple TLS in a Dicke ("super-qubit") state [35].

## 8   Data availability statement

All data needed to evaluate the conclusions in the paper are present in the paper.

## 9  Author Contributions

A.G., A.Y., J.S., and D.R. conceived the concept. B.Z. and D.R. performed the theoretical derivations and prepared the figures and videos. B.Z., D.R., and A.G. wrote the paper with contributions from R.I. and A.F. All authors reviewed and discussed the manuscript and made significant contributions to it.

## 10  Funding

This work was supported by the Israel Science Foundation (ISF) under Grant No. 00010001000, the National Natural Science Foundation of China under Grant No. 12104068, the Natural Science Foundation of Chongqing under Grant No. cstc2021jcyj-msxmX0684. D.R. acknowledges support by the PBC program of the Israel Council of Higher Education.

## 11  Acknowledgments

We acknowledge helpful communications with O. Kfir, Y. M. Pan and F. J. Garcia de Abajo.

## 12  Competing interests

The authors declare that they have no competing interests.